\documentclass[final,aps,prb,twocolumn,superscriptaddress,showpacs]{revtex4-1}

\usepackage{graphicx}
\usepackage{dcolumn}
\usepackage{bm}
\usepackage{xr}
\usepackage{soul,xcolor}
\setstcolor{red} 
\usepackage{kotex}

\usepackage{xr}

\makeatletter
\newcommand*{\addFileDependency}[1]{
  \typeout{(#1)}
  \@addtofilelist{#1}
  \IfFileExists{#1}{}{\typeout{No file #1.}}
}
\makeatother

\newcommand*{\myexternaldocument}[1]{%
    \externaldocument{#1}%
    \addFileDependency{#1.tex}%
    \addFileDependency{#1.aux}%
}

\myexternaldocument{SYAn_thru-hole_control_by_oxidation_ver21_SI} 

\begin{document}

\preprint{}

\title[]{Controlling GaN nucleation via O\textsubscript{2}-plasma-perforated graphene masks on c-plane sapphire}

\author{Su Young An}
\affiliation{Department of Physics and Research Institute for Basic Sciences, Kyung Hee University, Seoul, 02447, Republic of Korea}

\author{Chinkyo Kim}
\email{ckim@khu.ac.kr}
\affiliation{Department of Physics and Research Institute for Basic Sciences, Kyung Hee University, Seoul, 02447, Republic of Korea}
\affiliation{Department of Information Display, Kyung Hee University, Seoul, 02447, Republic of Korea}

\date{\today}

\begin{abstract}
Atomically thin, perforated graphene on $c$-plane sapphire functions as a nanoscale mask that enables GaN growth through thru-holes. We tune the perforated-area fraction $f_p$ by controlled O$_2$-plasma exposure and quantify its impact on early-stage nucleation: the nucleation-site density scales with $f_p$, while the nucleation-delay time decreases approximately as $1/f_p$. Time-resolved areal coverage and domain counts exhibit systematic $f_p$-dependent trends. A kinetic Monte Carlo (kMC) model that coarse-grains atomistic events—adatom arrival, surface diffusion, attachment at exposed sapphire within perforations, and coalescence (the first front–front contact between laterally growing domains)—reproduces these trends using a constant per-site nucleation rate. Fitting the kMC simulation data yields onset times \(t_{0}\) for the nucleation delay that closely match independently observed no-growth thresholds (Set~1: 28.5\,s vs \(\sim\)30\,s; Set~2: 38\,s vs \(\sim\)35\,s), validating the kMC–experiment mapping and highlighting plasma dose as an activation threshold for plasma-induced through-hole formation in 2D materials. Together, experiment and kMC identify $f_p$ as a single, surface-engineerable parameter governing GaN nucleation statistics on perforated graphene masks, providing a quantitative basis and process window for epitaxial lateral overgrowth (ELOG)/thru-hole epitaxy (THE) workflows that employ two-dimensional masks.
\end{abstract}



\maketitle

\section{Introduction}
Heteroepitaxial GaN on $c$-plane sapphire (Al$_2$O$_3$) suffers from large lattice and thermal-expansion mismatch, yielding high threading-dislocation (TD) densities that degrade device performance.\cite{Amano-APL-48-353,Mathis-JCG-231-371,Moram-JAP-106-073513,Kaganer-PRB-72-045423}
Epitaxial lateral overgrowth (ELOG) on patterned dielectrics is an established route to mitigate TDs,\cite{Kim-CGD-20-6198} but nanoscale mask patterning by nanoimprint or electron-beam lithography, while offering high-resolution selectivity,\cite{Kupers-SST-32-115003,Lin-MM-8-13} remains challenging to scale at wafer level. This motivates mask concepts that achieve nanoscale selectivity without lithographic overhead---the direction pursued here.  Accordingly, we used SiO$_2$-patterned sapphire with identical circular windows (4\,$\mu$m, hexagonal array) to confine nucleation locally, which enabled time-resolved, window-by-window quantification of areal coverage and domain counts via straightforward image segmentation, and provided numerous statistical replicates across plasma conditions.

In recent years, thru-hole epitaxy (THE) has emerged as a promising strategy for crystallographically aligned and, when desired, detachable GaN domains by leveraging nanoscale openings in two-dimensional (2D) masks such as graphene and h-BN.\cite{Lee-CGD-22-6995,Jang-AMI-10-2201406,Lee-AEM-26-2301654}
In THE, nucleation occurs directly on the exposed substrate through these openings, followed by lateral overgrowth over the mask.
This mechanism is less constrained than remote epitaxy, which typically requires pristine monolayer 2D layers and strict control of polarity and thickness.
Related ideas have also been extended to solution-processed masks, where percolative transport through disordered or spin-coated multilayer flakes enables GaN nucleation at selective sites beneath the 2D material.\cite{Beak-arXiv,Ha-arXiv}
Taken together, these results motivate comparison with approaches that rely on intact 2D interlayers (remote or vdW epitaxy), summarized next.

A recent study established mechanistic principles for 2D-material–assisted nitride epitaxy: on single-crystal substrates, a graphene interlayer can relay the substrate potential to enable remote epitaxy, whereas on amorphous supports the process is van der Waals (vdW) in nature; notably, GaN on graphene/glass tends to be polycrystalline, while WS$_2$ can mediate single-crystal GaN.\cite{Chen-AM-35-2211075}
Specifically, MOVPE on graphene-coated $c$-plane sapphire realized \emph{remote epitaxy} of GaN by employing an O$_2$-plasma pretreatment to increase nucleation sites on graphene and a pulsed NH$_3$ scheme that preserves the graphene interlayer.\cite{Lee-NL-23-11578}
Beyond sapphire, single-crystalline GaN has also been realized on CMOS-compatible Si(100) using a single-crystal graphene/SiO$_2$ interlayer; NH$_3$ pretreatment induces C--N bonding on graphene that triggers nitride nucleation and yields an in-plane–aligned, continuous single-crystal film.\cite{Feng-AFM-29-1905056}
Patterned graphene masks have likewise been used for GaN ELOG on GaN/sapphire templates, where micrometer-scale windows enabled lateral overgrowth with reduced TD density.\cite{Tao-JJAP-63-025503}
In contrast, here we \emph{perforate} graphene with O$_2$ plasma to define nanoscale \emph{thru-holes} that expose the underlying $c$-plane sapphire, enabling nucleation on the substrate (THE).

A key outstanding challenge is to precisely and reproducibly control the density and distribution of these nanoscale openings in a scalable manner.
In this work, we address this gap by tuning the \emph{perforated-area fraction} of graphene, denoted $f_p$, via O$_2$ plasma treatment, which introduces nanoscale vacancies/holes in the graphene lattice.\cite{Liu-NL-8-1965}
Unlike prior studies that relied on uncontrolled defects or transfer-induced openings, our approach enables systematic, sub-lithographic control of perforation and, consequently, of GaN nucleation behavior. By varying the plasma exposure time, we reliably tune the perforated morphology in graphene films directly grown on $c$-plane sapphire by CVD, and we quantify its impact on nucleation using an \emph{effective} perforated-area fraction $f_p^{\mathrm{eff}}$ inferred from model–experiment comparison when direct AFM quantification is not available.

Through GaN growth experiments and a kinetic Monte Carlo (kMC) model that coarse-grains atomistic events (arrival, diffusion, attachment at exposed sapphire within perforations, and coalescence) into rates, we demonstrate that both the \emph{fraction} and \emph{distribution} of perforated area strongly influence the spatial and temporal evolution of GaN domain nucleation.
In particular, the nucleation-site density scales with $f_p$, while the nucleation-delay time decreases approximately as $1/f_p$.  These findings highlight O$_2$ plasma–engineered graphene masks as a scalable platform for controlled nucleation in GaN epitaxy and provide a quantitative basis for optimizing ELOG/THE workflows with 2D masks, using $f_p^{\mathrm{eff}}$ as a practical surface-engineering parameter.

\section{Methods}

\subsection{Experimental methods}

A nominal 50\,nm-thick SiO$_2$ film was deposited on $c$-plane sapphire (Al$_2$O$_3$) and patterned with circular openings via photolithography. The circular openings were arranged in a hexagonal close-packed layout (diameter $\sim$4\,$\mu$m). We used SiO$_2$-patterned substrates, rather than bare sapphire, to confine GaN nucleation to isolated and identical windows. This geometry enabled time-resolved, window-by-window quantification of (i) areal domain coverage and (ii) domain counts with straightforward image segmentation, and it provided many statistical replicates per sample. As a result, we could monitor the time evolution of coverage and domain number with high fidelity when comparing different O$_2$ plasma etching conditions.

Graphene was synthesized \emph{directly on the patterned sapphire} by chemical vapor deposition (CVD) at 1050$^\circ$C using CH$_4$ (18\,sccm) as the carbon source, H$_2$ (10\,sccm) as a co-reactant/reducing gas, and Ar as the carrier. Substrates were first annealed for 15\,min in Ar (780\,Torr, 0.3\,slm), followed by 10\,min growth and two-step cooling by retraction of the loading arm.

Thick graphene films were intentionally grown so that, even after O$_2$ plasma treatment, the remaining graphene retained sufficient integrity to function as a selective growth mask. To this end, the Ar flow during CVD was minimized to enable rapid thickening. In contrast, monolayer-style recipes typically require longer thermal budgets to accumulate comparable carbon throughput, which may undesirably modify the sapphire surface and affect GaN nucleation. The thick-graphene condition reduced such thermal exposure and supported more controlled GaN epitaxy.

O$_2$ plasma etching was performed in an oxygen asher to introduce nanoscale perforations in graphene. To assess whether plasma parameters provide meaningful control over GaN domain nucleation, we employed two sets of etching conditions:

\begin{itemize}
  \item \textbf{Set 1:} RF power 20\,W, pressure 50\,Torr; etch times 30, 35, 40, 50, and 60\,s.
  \item \textbf{Set 2:} RF power 17\,W, pressure 100\,Torr; etch times 40, 50, 60, 80, and 100\,s.
\end{itemize}
The degree of perforation was controlled by plasma exposure time (longer times $\Rightarrow$ higher perforation). Direct quantification of the perforated-area fraction was feasible only for a subset of samples; for analysis and modeling, we therefore use an \emph{effective} perforated-area fraction $f_p^{\mathrm{eff}}$ inferred as described below.

GaN growth was carried out on the etched graphene/$c$-plane sapphire using hydride vapor phase epitaxy (HVPE) at 942$^\circ$C with N$_2$ as the carrier gas. HCl (9\,sccm) and NH$_3$ (0.6\,slm) served as the precursors via:
\begin{equation}
2\,\mathrm{Ga} + 2\,\mathrm{HCl} \rightarrow 2\,\mathrm{GaCl} + \mathrm{H}_2,
\end{equation}
\begin{equation}
\mathrm{GaCl} + \mathrm{NH}_3 \rightarrow \mathrm{GaN} + \mathrm{HCl} + \mathrm{H}_2.
\end{equation}
For each plasma condition, the GaN growth duration was varied to resolve early-time nucleation and coverage evolution.  For simulation–experiment comparison, one routine step was taken as a fixed time increment $\Delta$t, and the step index was linearly mapped to growth duration.

\subsection{kMC model}

We implemented a discrete-time kMC on a square lattice of size $L\times L$ ($N_t=L^2$; typically $L=100$). Boundaries are open (non-periodic), so edge cells have fewer neighbors than interior cells. All sites are initially empty.  Each time step (one ``routine'') consists of two stages: nucleation and growth.

\subsubsection{Nucleation}
We perform $N_o$ nucleation \emph{trials} at uniformly random lattice sites (sampling with replacement). For each trial, if the chosen site is currently empty, it nucleates with probability $P_0$ (constant across steps). Here, the per-site nucleation rate \(\lambda\) denotes the instantaneous event rate for a candidate opening that has not yet nucleated; in the discrete-time kMC we use a per-step probability \(P_0\) (with \(\lambda \approx P_0/\Delta t\) for \(P_0\!\ll\!1\)).  Thus, an empty site experiences an \emph{effective} per-step probability of nucleation \(P_{\mathrm{eff}}\) (conditional on not yet nucleated), equivalently a per-site rate \(\lambda_{\mathrm{eff}}\approx P_{\mathrm{eff}}/\Delta t\) for \(P_{\mathrm{eff}}\!\ll\!1\).

\[
P_{\mathrm{eff}} \;=\; 1-\bigl(1-\frac{P_0}{N_t}\bigr)^{N_o} \;\approx\; \frac{N_o}{N_t}\,P_0 
\quad (P_0\ll 1),
\]
which corresponds to a continuous-time rate $r_0 \approx (N_o/N_t)\,P_0/\Delta t$ when one step is taken as $\Delta t$.
After nucleation, a site is considered occupied and no longer eligible to nucleate.

\subsubsection{Growth and coalescence}
Following nucleation, occupied sites expand by one lattice cell to their four nearest neighbors (Manhattan metric) in a parallel update, producing diamond-like fronts. When expanding fronts meet, clusters merge; connected components are defined with a 4-neighbor connectivity rule (the same rule used for counting).

\subsubsection{Observables and averaging}
At each step we record (a) the areal coverage $\theta$ as the fraction of occupied sites and (b) the number of connected domains (4-neighbor components). Time is reported as $t=n_{\mathrm{step}}\Delta t$; for experiment–simulation overlays, the step index is linearly mapped to growth duration. All curves are ensemble-averaged over $N_a$ independent realizations.

\subsubsection{Estimating and mapping the perforation fraction}
Because, in the rare-event limit, the effective per-step nucleation probability for an empty site scales as
\(P_{\mathrm{eff}}\approx (N_o/N_t)\,P_0\), we model the impact of perforation by choosing $N_o \propto f_p^{\mathrm{eff}}\,N_t$ for fixed $P_0$ and $\Delta t$. Here $f_p^{\mathrm{eff}}$ denotes an \emph{effective} perforated-area fraction obtained by inverse modeling: for each plasma condition, $f_p^{\mathrm{eff}}$ is selected to best reproduce the experimentally measured time series of (i) areal coverage and (ii) domain counts under a single set of $P_0$, $\Delta t$ and counting rules. Where AFM/SEM segmentation was reliable, those samples were used to anchor the monotonic $f_p^{\mathrm{eff}}(t_{\mathrm{etch}})$ mapping; otherwise, $f_p^{\mathrm{eff}}$ was inferred solely from the time-series fit. Throughout the paper, $f_p$ refers to $f_p^{\mathrm{eff}}$ unless explicitly stated. 

For each etch condition (fixed power/pressure), the kMC holds \(N_o\)---and thus \(f_p\propto N_o/N_t\)---constant within a run (time‐independent mask); the dependence on etch time \(t_{\mathrm e}\) is encoded via the fitted nucleation delay \(\tau_{\mathrm{delay}}(t_{\mathrm e})\), which is equivalent to rescaling the effective per-site rate \(\propto f_p^{\mathrm{eff}}(t_{\mathrm e})\) near onset.

Additional Set~2 datasets are provided in the Supplementary Information.

\section{Results and discussion}

\begin{figure}
\includegraphics[width=1.0\columnwidth]{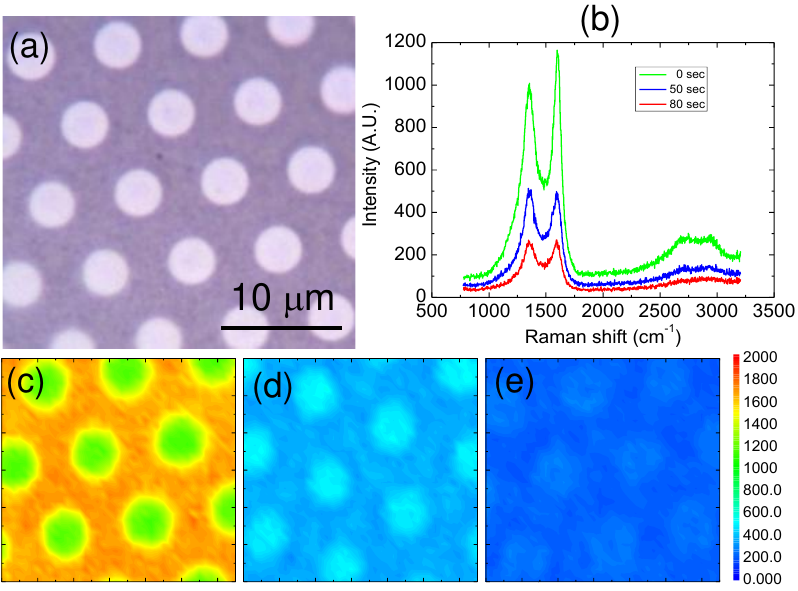}
\caption{(a) Optical micrograph of graphene grown on a SiO$_2$-patterned $c$-plane sapphire substrate. 
(b) Raman spectra acquired at the circular opening regions after different O$_2$-plasma durations. 
Raman maps of the G-peak intensity, $I_\mathrm{G}$, of graphene after (c) 0\,s (no etch), (d) 50\,s, and (e) 80\,s O$_2$-plasma exposure. 
The openings form a hexagonal array with \emph{nominal diameter} $\sim 4~\mu$m and center-to-center spacing $\sim 8~\mu$m.}
\label{Raman_O2_plasma_effect_on_graphene_before_growth}
\end{figure}

Unless noted otherwise, we report trends versus O$_2$-plasma exposure time; where $f_p$ is used, it denotes the effective fraction $f_p^{\mathrm{eff}}$ obtained from the inverse modeling procedure described in Methods.

\subsection{O$_2$ plasma treatment for perforating graphene}

O$_2$ plasma progressively removes carbon from graphene; by tuning dose (power$\times$time), we induce partial removal within the circular windows, i.e., perforation. 
Figure~\ref{Raman_O2_plasma_effect_on_graphene_before_growth} shows Raman maps of the G-peak intensity, $I_\mathrm{G}$, of graphene directly grown on SiO$_2$-patterned $c$-plane sapphire prior to GaN growth. 
In the as-grown state, each circular opening is covered by multilayer graphene. 
With increasing O$_2$-plasma duration, a reduction of $I_\mathrm{G}$ within openings is clearly observed, consistent with partial etching of graphene. 
Up to at least 50\,s, a finite G peak remains detectable inside the openings, indicating that a portion of graphene persists after moderate etching.

\begin{figure}
\includegraphics[width=1.0\columnwidth]{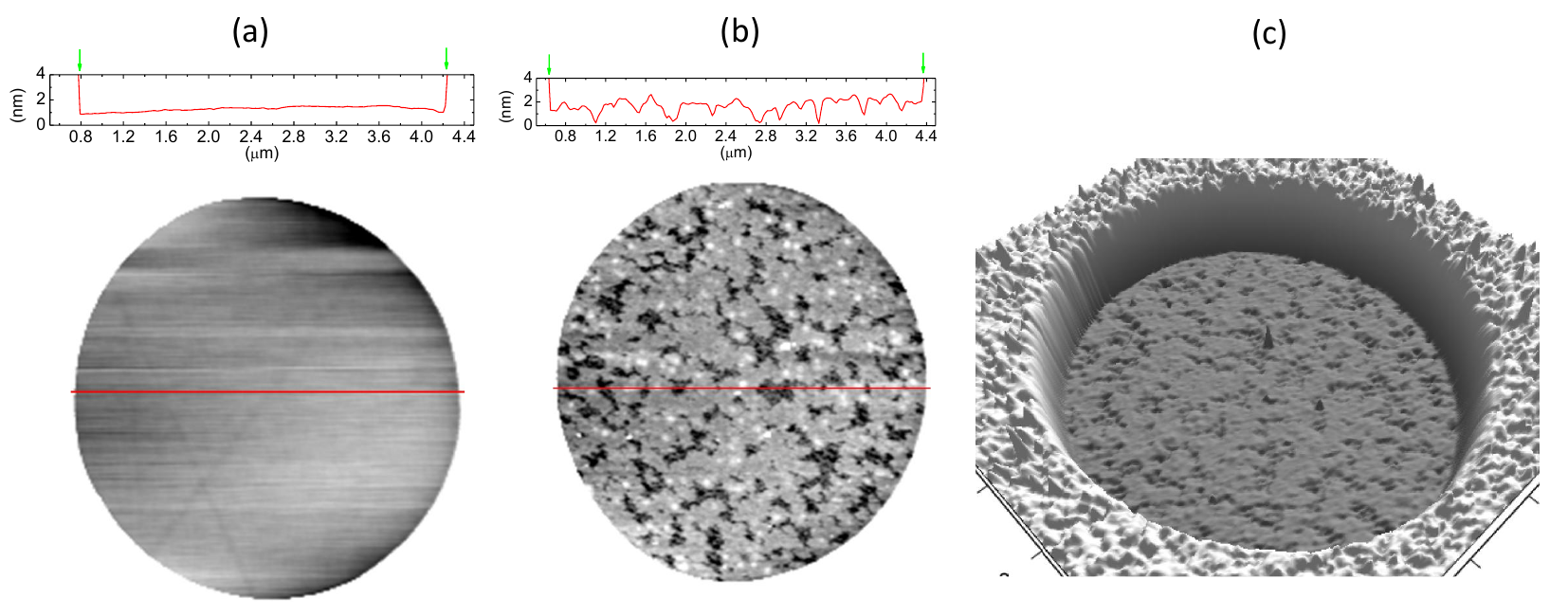}
\caption{AFM topography and height profiles along the red line for (a) a bare sapphire region (no graphene) and (b) graphene after 35\,s of O$_2$ plasma within a circular opening defined by the SiO$_2$ mask. 
Perforated (dark) features appear inside the circular scan region in (b); segmentation yields a perforation fraction $f_p \approx 0.14$. 
(c) 3D rendering of the AFM image in (b). 
Green arrows in (a) and (b) mark the SiO$_2$ mask edges.}
\label{AFM_O2_plasma_effect_on_graphene}
\end{figure}

To assess perforation morphology, AFM topography was measured on as-grown and O$_2$-plasma–treated samples (Fig.~\ref{AFM_O2_plasma_effect_on_graphene}). 
The as-grown graphene exhibits a contiguous, thick layered film across the opening area. 
After a 35\,s O$_2$-plasma treatment, numerous dark depressions appear within the circular region, corresponding to exposed sapphire; for this condition, segmentation of the AFM image in Fig.~\ref{AFM_O2_plasma_effect_on_graphene}(b) gives $f_p \approx 0.14$. 
Individual perforations expose the sapphire surface, with lateral dimensions of sub-100~nm and depths of a few nanometers.  Consistent with this interpretation, the next subsection shows that GaN nucleates within these dark regions under HVPE growth, confirming exposure of the substrate. 
With further O$_2$-plasma increase, adjacent perforations coalesce and large areas of thinner graphene are removed, leaving only thicker graphene domains.

\subsection{Effect of O$_2$ plasma treatment on the nucleation and growth of GaN domains}

We examine how O$_2$ plasma applied to graphene-covered $c$-plane sapphire alters GaN nucleation and lateral growth. O$_2$ plasma breaks C--C bonds in graphene,\cite{Childres-NJP-13-025008} producing perforations whose presence is evident in AFM topography and whose partial carbon removal is supported by Raman spectroscopy. Perforations expose local regions of the $c$-plane sapphire, increasing the likelihood that GaN nucleates on those sites once thinner graphene is fully opened.

\begin{figure}
\includegraphics[width=1.0\columnwidth]{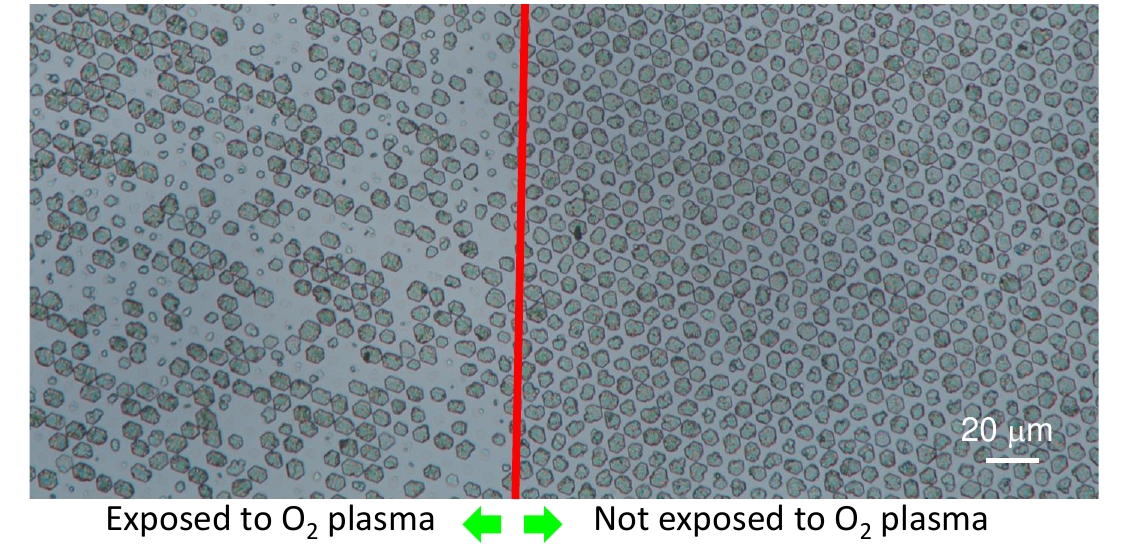}
\caption{Optical microscopy of GaN domains grown on SiO$_2$-patterned $c$-plane sapphire with and without prior O$_2$ plasma on the \emph{uncovered} regions (slide-glass masking; 20~W, 1~min). Plasma-treated bare sapphire shows reduced nucleus count and areal coverage, evidencing a nucleation-unfavorable surface modification.}
\label{O2-plasma-effect}
\end{figure}

Beyond creating openings, O$_2$ plasma can also modify the bare $c$-plane sapphire surface itself.\cite{Zhang-ASS-285-211} Zhang \emph{et~al.}\cite{Zhang-ASS-285-211} reported terrace etching, step roughening, and pit formation depending on dose, all of which can influence subsequent nucleation. To isolate this effect, we exposed \emph{graphene-free} SiO$_2$-patterned sapphire to O$_2$ plasma using a slide-glass mask (20~W, 1~min). As shown in Fig.~\ref{O2-plasma-effect}, plasma-treated bare sapphire exhibits fewer nuclei and lower areal coverage than untreated regions, indicating a surface modification that is unfavorable for GaN nucleation.

Taken together, O$_2$ plasma exerts two opposing influences: (i) a \emph{nucleation-favorable} effect by creating perforations in graphene that expose sapphire, and (ii) a \emph{nucleation-unfavorable} effect by degrading the bare sapphire surface. At short plasma durations, as perforations begin to form and remain isolated, the favorable effect dominates and crystallographically aligned GaN nuclei appear readily. At higher doses, neighboring openings merge and the underlying sapphire experiences stronger plasma exposure; the unfavorable effect becomes prominent and the nucleus count decreases relative to fresh sapphire. The net trends therefore reflect the balance of these two effects.

\begin{figure}
\includegraphics[width=1.0\columnwidth]{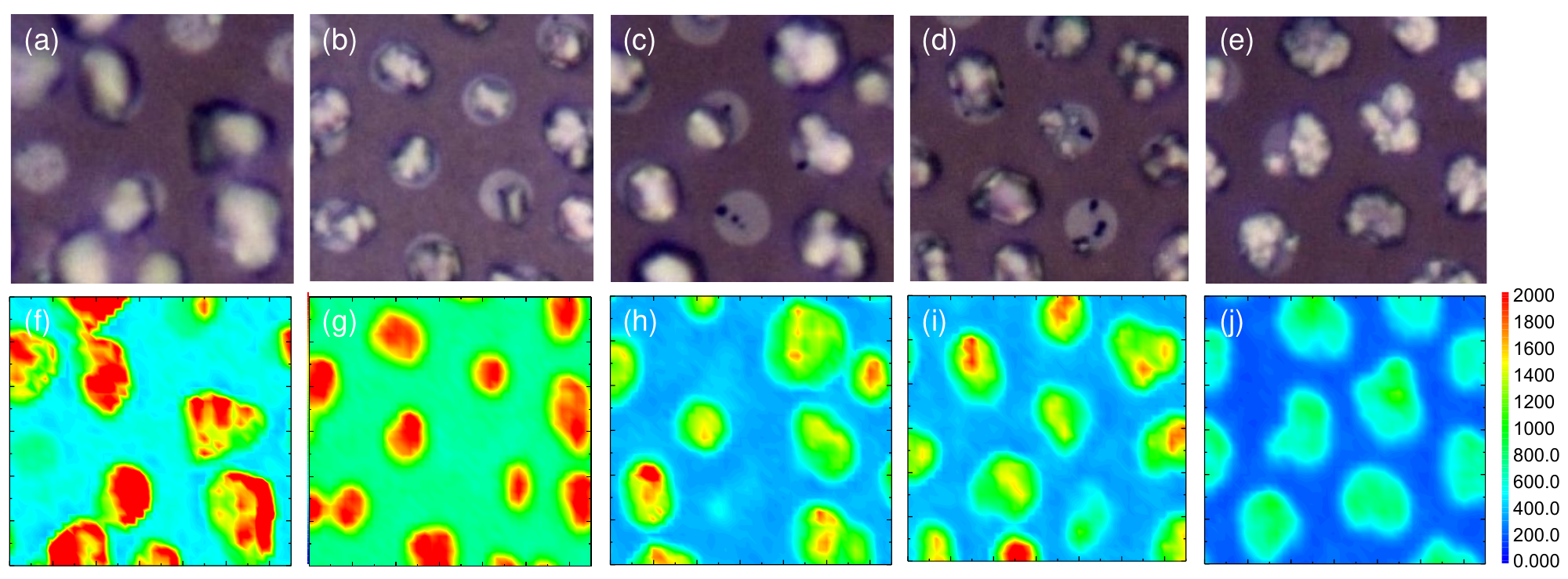}
\caption{Optical microscopy images for samples where GaN was grown on graphene/$c$-plane sapphire after O$_2$-plasma treatments of (a) 40\,s, (b) 50\,s, (c) 60\,s, (d) 80\,s, and (e) 100\,s. (f–j) Corresponding Raman $I_\mathrm{G}$ (G-peak) intensity maps for the same samples. A finite G peak persists, indicating that graphene remains after high-temperature growth of GaN.}
\label{Raman-map-after-GaN-growth}
\end{figure}

The integrity of the graphene interlayer during high‐temperature GaN growth is critical. Post‐growth Raman mapping (Fig.~\ref{Raman-map-after-GaN-growth}) shows a persistent G peak after growth at $\sim$1000$^\circ$C; with increasing O$_2$‐plasma duration the G‐peak intensity decreases consistently, indicating partial thinning rather than removal. Moreover, G‐peak intensity is concentrated where GaN domains are present. This pattern suggests thermal/oxidative degradation of \emph{exposed} graphene in the HVPE environment. Two scenarios are plausible: degradation during GaN growth or during the subsequent unloading/cooling stage. If substantial oxidation occurred \emph{during} growth, the perforated fraction would vary dynamically, weakening the observed dependence on etch duration and causing noticeable disagreement with the kMC trends shown later.  Thus, this observation indicates that in‐growth degradation is minimal. A consistent interpretation is that most graphene oxidation occurs \emph{after} growth, i.e., after perforations have already been filled, so GaN shields the underlying graphene and a strong G peak persists beneath GaN domains. This supports modeling with a time‐independent $f_p$ during growth.

\begin{figure}
\includegraphics[width=1.0\columnwidth]{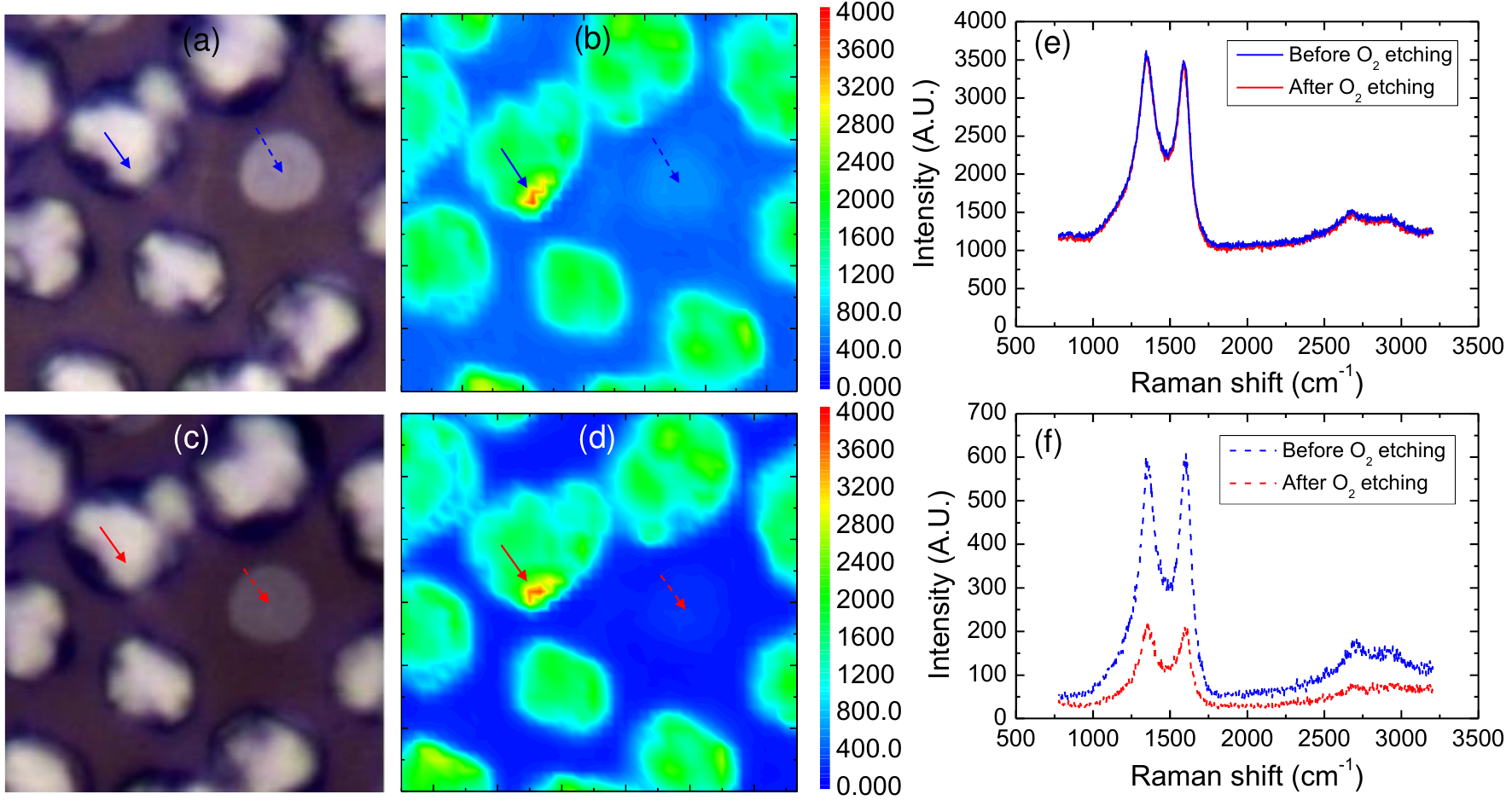}
\caption{Before and after O$_2$-plasma etching of GaN grown on graphene/SiO$_2$/sapphire. (a,c) Optical microscopy images before and after etching. (b,d) Corresponding Raman G-peak ($I_G$) maps. (e) Raman spectra at the solid-arrow positions in (b) and (d) (beneath GaN) show strong G peaks without intensity loss upon etching, whereas (f) spectra at the dashed-arrow positions (no GaN within the circular opening) show reduced G intensity after etching. Because O$_2$ plasma damages exposed graphene but does not etch through GaN, the persistence of the G peak beneath GaN confirms that graphene remains at the GaN/sapphire interface.}
\label{Interfacial-graphene}
\end{figure}

To test whether graphene remains interfacial, we performed post‐growth O$_2$‐plasma etching. As shown in Fig.~\ref{Interfacial-graphene}, graphene beneath GaN survives the etch whereas graphene in uncovered regions within the circular openings is diminished, confirming that graphene resides at the GaN/sapphire interface. This behavior contrasts with the previously reported “lift‐up” of spin‐coated 2D masks\cite{Ha-arXiv} and is consistent with GaN overgrowing a pre‐existing graphene mask. 

Relating the density of GaN nuclei directly to the density of graphene openings is nontrivial. If growth proceeds long enough for lateral overgrowth to fully cover a $4~\mu$m-diameter opening, initially distinct nuclei coalesce and the original nucleation density is no longer observable. Conversely, at very short growth times, not all available perforation sites will have nucleated due to the stochastic nature of nucleation. Thus simple nucleus counting cannot, by itself, recover the perforation density. Instead, we (1) measured the time evolution of the domain count and the areal coverage within each SiO$_2$-defined opening, and (2) analyzed these data using a kMC model based on classical nucleation theory. This approach enables inference of the relation between perforation and nucleation statistics.

\begin{figure}
\includegraphics[width=0.9\columnwidth]{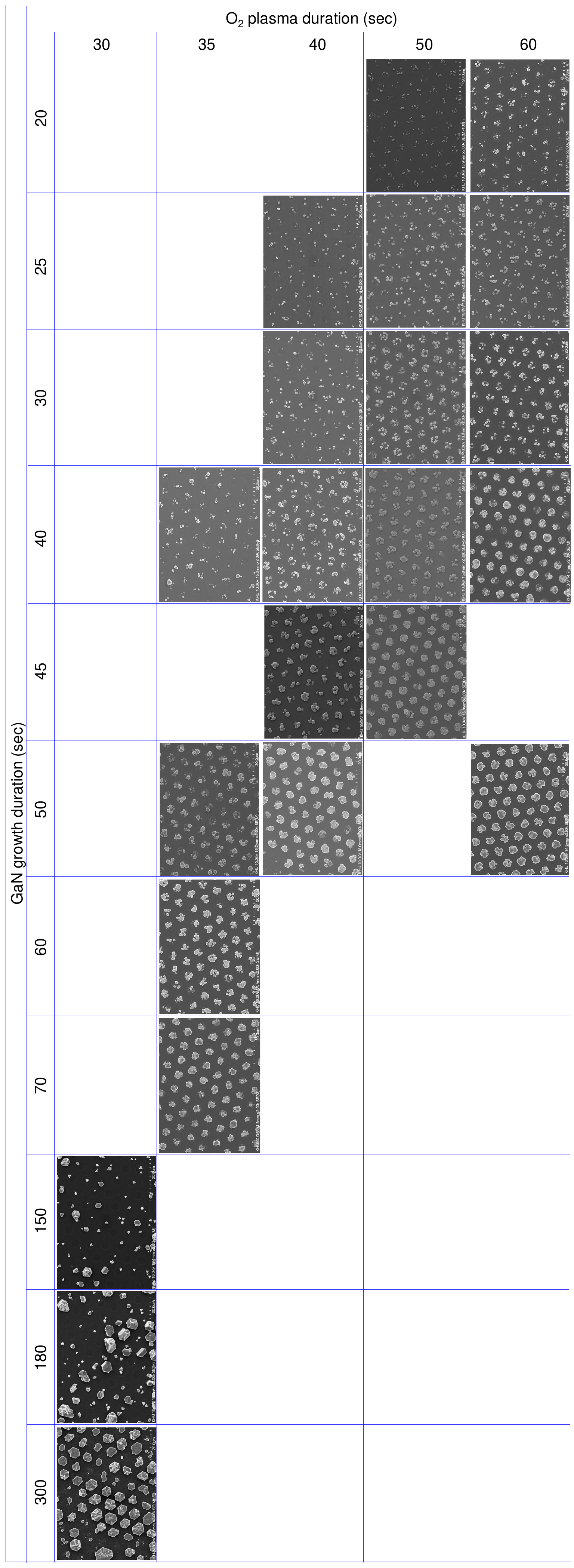}
\caption{SEM images of GaN domains grown on graphene/$c$-plane sapphire for various O$_2$ plasma conditions and growth durations. For a fixed plasma condition, domains within each $4~\mu$m-diameter opening expand and merge with increasing growth time.}
\label{SEM_GaN_on_O2-etched_graphene}
\end{figure}

\begin{figure}
\includegraphics[width=1.0\columnwidth]{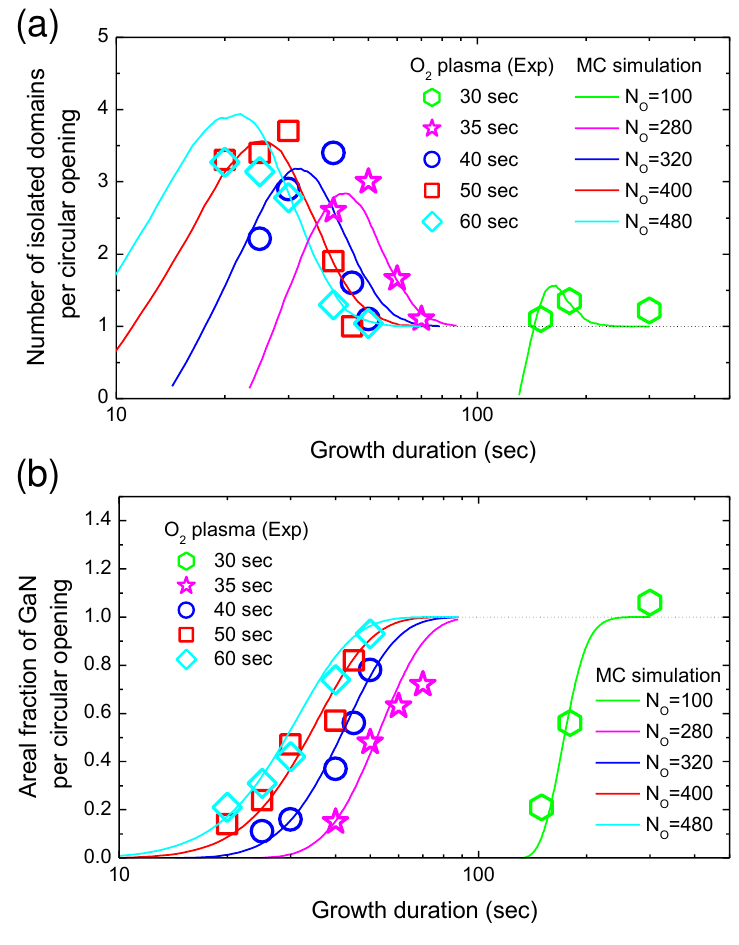}
\caption{Time evolution of domain count and areal coverage for Set~1. Experimental symbols are averages over multiple openings per sample. The simulation curves use a kMC model with a constant per-step nucleation probability (see Methods):  kMC parameters: $N_{\rm t}=1\times10^4$, $P_0=5\times10^{-4}$, $N_{\rm r}=100$--$300$, $N_{\rm a}=2000$.}
\label{Time-evolution-of-GaN-domains}
\end{figure}

For quantitative analysis, we evaluated (i) the number of isolated GaN domains and (ii) the areal coverage within several dozen openings per sample based on SEM images of each sample shown in fig.~\ref{SEM_GaN_on_O2-etched_graphene}. To ensure consistency across samples, features smaller than $0.2~\mu\mathrm{m}^2$ were excluded; such small features can include partially reacted material or byproducts (e.g., GaCl residues) lacking clear crystalline morphology. Our goal is to compare relative nucleation behavior across conditions rather than to quantify absolute mass. Applying a uniform exclusion criterion yields an internally consistent comparison focused on stable epitaxial domains. Counts were obtained in ImageJ and averaged across openings; the experimental symbols in Fig.~\ref{Time-evolution-of-GaN-domains} plot the resulting time series.

\subsection{kMC simulation of nucleation and growth on O$_2$-plasma–treated graphene}

For the 35\,s O$_2$-plasma condition [Fig.~\ref{AFM_O2_plasma_effect_on_graphene}(b)], AFM topography yields a perforation fraction $f_p \approx 0.14$, which we use to anchor the monotonic mapping between plasma exposure time and the effective perforation fraction $f_p^{\mathrm{eff}}$ employed in the kMC analysis.

\begin{figure}
\includegraphics[width=1.0\columnwidth]{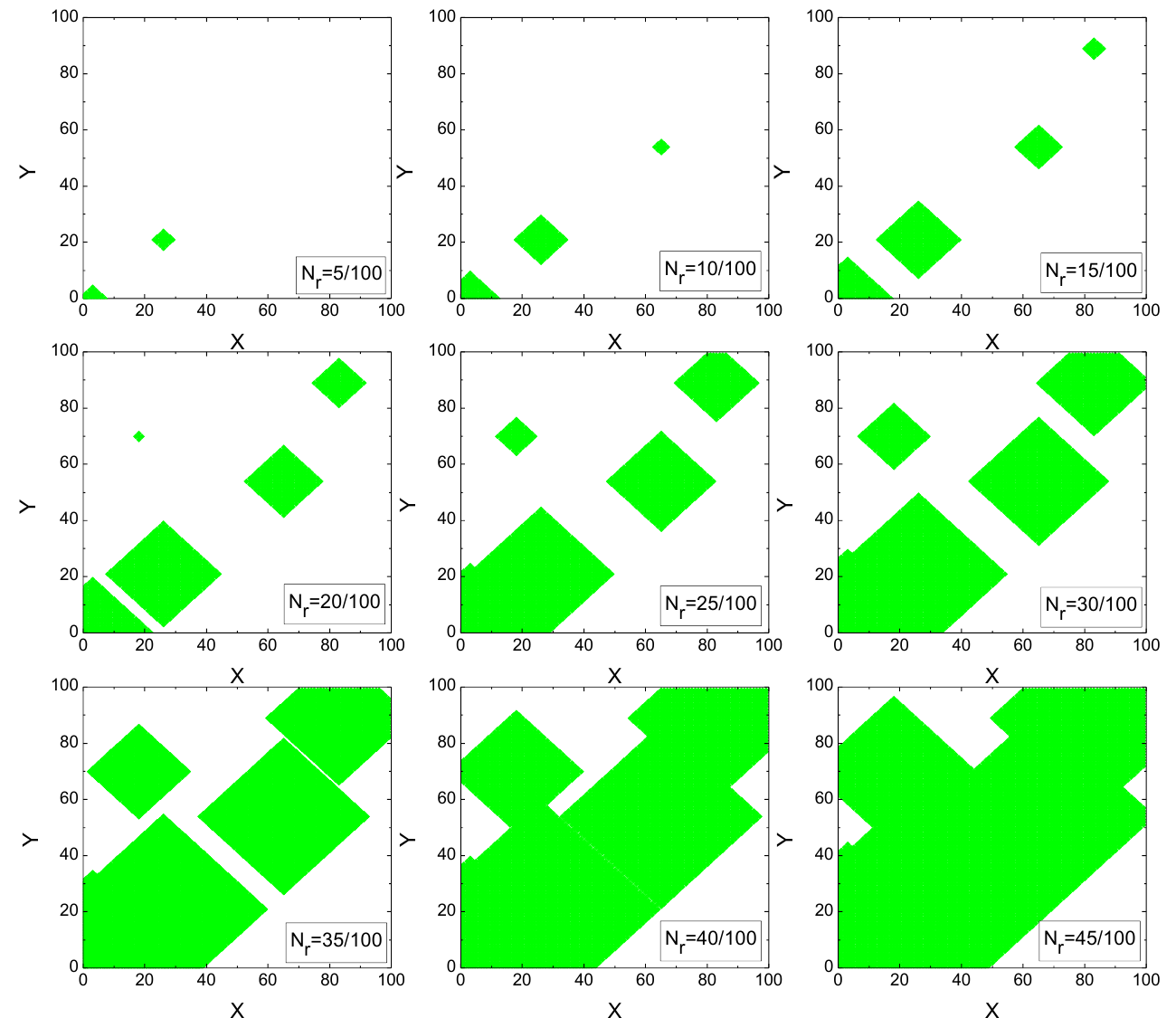}
\caption{Exemplary time evolution of domains simulated by kMC.}
\label{Simulation}
\end{figure}

To quantify the observations, we implemented a discrete-time kMC on a $100\times100$ lattice (open boundaries, 4-neighbor connectivity). Each step applies $N_o$ random nucleation trials over the lattice with constant per-trial probability $P_0=5\times10^{-4}$ (rare-event limit), followed by one-pixel 4-neighbor growth/coalescence; one step is a fixed time increment $\Delta t$, so $t=j\,\Delta t$.  In the rare-event limit, the effective per-step per-site probability of nucleation scales as
\(P_{\mathrm{eff}} \approx (N_o/N_t)\,P_0\), so we model perforation by $N_o \propto f_p^{\mathrm{eff}} N_t$ (no explicit spatial mask). For each plasma condition we simulated $N_{\rm r}=100$–$300$ steps and averaged $N_{\rm a}=2000$ realizations. Figure~\ref{Simulation} shows the exemplary time evolution of individual domains obtained by kMC simulation.

The kMC outputs—(1) ensemble-averaged number of isolated domains (4-neighbor components) and (2) areal coverage within each $4~\mu$m-diameter opening—are plotted as curves in Fig.~\ref{Time-evolution-of-GaN-domains} alongside experimental symbols. The simulation curves use the constant‐rate kMC model (constant per‐step per‐site nucleation probability), and the fitted \(\tau_{\mathrm{delay}}\propto 1/f_p\) near onset is consistent with \(f_p\propto (t_{\mathrm e}-t_0)\).  For a fixed plasma condition, the domain count versus $t$ rises, saturates, and then decreases as islands coalesce, consistent with classical nucleation-and-growth kinetics. Two systematic trends appear and are captured by the model: (i) the \emph{peak time shifts earlier} with larger $f_p$ because the mean opening spacing $L \sim 1/\sqrt{f_p}$ is smaller, so coalescence $t_{\mathrm{coal}}\!\sim\!L$ is reached sooner; (ii) the \emph{peak height decreases} as $f_p$ decreases because the seed-injection rate per area ($\propto f_p$) is lower and each seed at small $f_p$ draws adatoms from a larger capture zone, reducing the number of simultaneously distinct islands. The areal coverage shows a steeper early-time rise at smaller $f_p$, consistent with larger capture zones that sweep area before encountering neighbors; higher $f_p$ produces earlier encounters that temper the net area added per step.

\begin{figure}
\includegraphics[width=0.97\columnwidth]{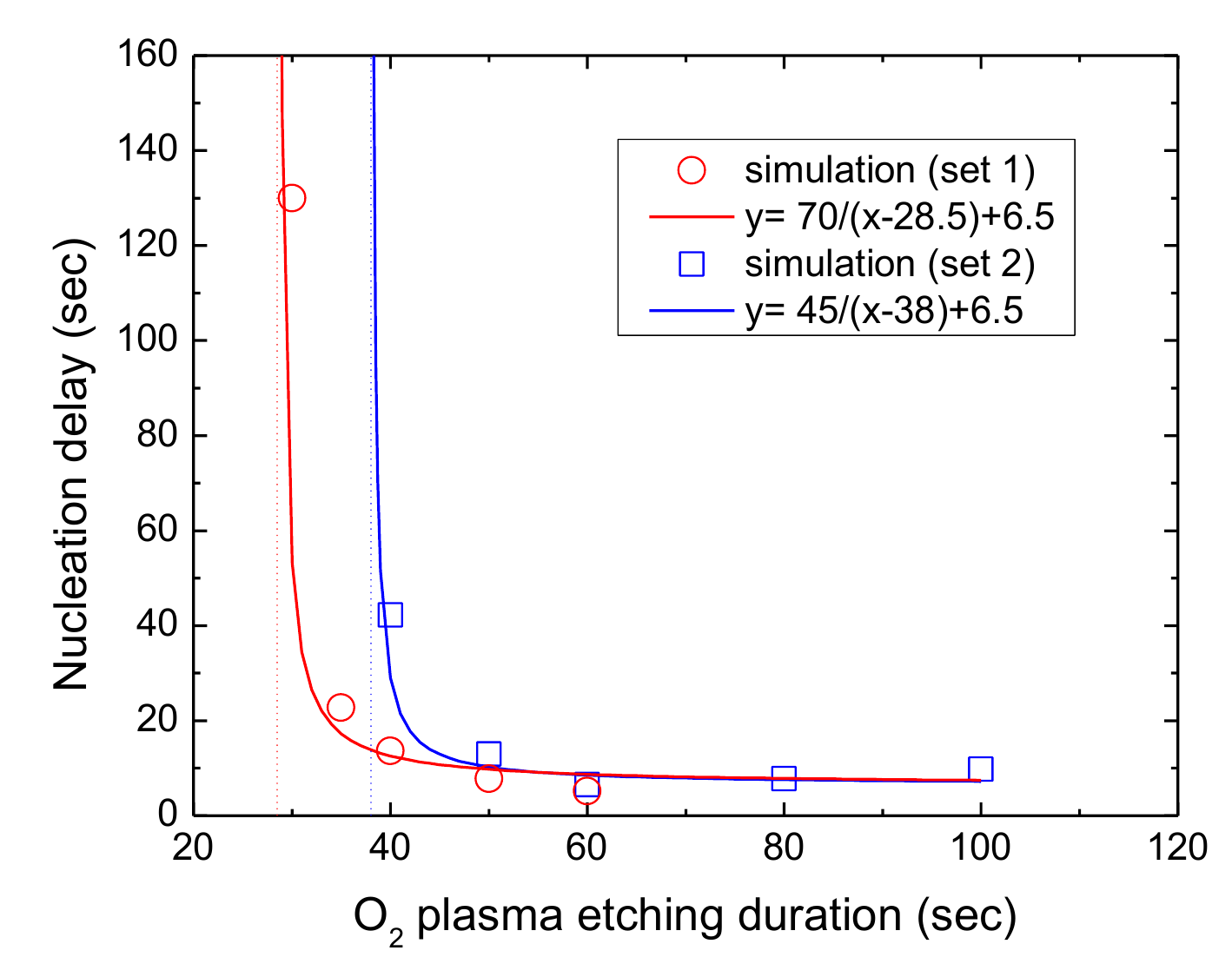}
\caption{Nucleation delay $\tau_{\mathrm{delay}}$ obtained by aligning simulation to experiment (coverage curves), plotted versus O$_2$-plasma etch time $t_{\mathrm{e}}$. The shifted-hyperbola fit
$\tau_{\mathrm{delay}}=\tau_0 + A/(t_{\mathrm{e}}-t_0)$
yields $A=70~\mathrm{s}^2$, $t_0=28.5~\mathrm{s}$, and $\tau_0=6.5~\mathrm{s}$. The fitted onset values \(t_0\) for the two sample sets are very close to their respective observed no-growth thresholds (Set~1: \(\sim\)30\,s; Set~2: \(\sim\)35\,s), consistent with the interpretation that \(\tau_{\mathrm{delay}}\!\to\!\infty\) when perforation is insufficient.}
\label{Nucleation-delay}
\end{figure}

The extracted delays versus etch time are summarized in Fig.~\ref{Nucleation-delay} and follow a shifted hyperbolic form
\[
\tau_{\mathrm{delay}}(t_{\mathrm e})=\tau_0+\frac{A}{t_{\mathrm e}-t_0},
\]
where \(t_{\mathrm e}\) is the O$_2$-plasma etch time; the best-fit parameters are \(A=70~\mathrm{s}^2\), \(t_0=28.5~\mathrm{s}\), and \(\tau_0=6.5~\mathrm{s}\).  The vertical asymptote at $t_{\mathrm{e}}=t_0$ represents the effective onset of perforation sufficient to expose sapphire. Notably, the onset times extracted from the kMC results closely reproduce the independently observed “no-growth” thresholds: Set~1 yields \(t_{0}=28.5~\mathrm{s}\) versus \(\sim\!30~\mathrm{s}\) observed, and Set~2 yields \(t_{0}=38~\mathrm{s}\) versus \(\sim\!35~\mathrm{s}\) (Fig.~\ref{Nucleation-delay}). This agreement supports the near-onset mapping \(f_p^{\mathrm{eff}}\!\propto\!(t_{\mathrm e}-t_{0})\) and validates using a time-independent mask within a growth run.

The physical origin of the inverse trend can be understood as follows: if the number of effective nucleation conduits (through-holes/activated openings) increases approximately linearly near onset, $N_{\mathrm{eff}}(t_{\mathrm{e}})\!\approx\! a\,(t_{\mathrm{e}}-t_0)$ for $t_{\mathrm{e}}\!>\!t_0$, and nucleation is Poisson with per-site attempt rate $k_0$, then the total rate is
$\Lambda(t_{\mathrm{e}})=k_0 N_{\mathrm{eff}}(t_{\mathrm{e}})$ and the mean waiting time scales as
\[
\tau_{\mathrm{delay}}(t_{\mathrm{e}})\approx \frac{1}{\Lambda(t_{\mathrm{e}})}=\frac{1}{k_0 a}\,\frac{1}{\,(t_{\mathrm{e}}-t_0)\,}.
\]
Allowing a finite detection floor $\tau_0$ gives the fitting form above. Equivalently, if the open-area fraction satisfies $f_p(t_{\mathrm{e}})\!\simeq\! \alpha\,(t_{\mathrm{e}}-t_0)$ in the early regime, then $\tau_{\mathrm{delay}}\!\propto\!1/f_p$, as observed.

Experimentally, the time-series trends and inverse-modeled \(f_p^{\mathrm{eff}}\) are consistent with this monotonic scaling: near onset \(f_p \!\propto\! (t_e-t_0)\), yielding \(\tau_{\mathrm{delay}}\!\propto\! 1/f_p\). However, because \(f_p\) was not independently measured for all samples and early-time counts are affected by coalescence/capture zones, we refrain from claiming strict linearity from raw counts alone.

Experimental data for Set~2 exhibit similar trends to those in the main text (Supplementary Fig.~S1 and Supplementary Fig.~S2); the corresponding nucleation delays are plotted in Fig.~\ref{Nucleation-delay}.

\section{Conclusion}
We demonstrated that multilayer graphene grown directly on $c$-plane sapphire can serve as a nanoscale mask for GaN thru-hole epitaxy/ELOG, with its perforation controllably tuned by O$_2$ plasma parameters (RF power, exposure time). Using an effective perforated-area fraction $f_p^{\mathrm{eff}}$ (anchored by AFM where available and inferred otherwise), we quantified how plasma dose regulates nucleation statistics: the nucleation-site density increases with $f_p^{\mathrm{eff}}$, while the nucleation-delay time decreases approximately as $1/(t_{\mathrm{e}}-t_0)$ (equivalently, $\propto 1/f_p$ in the early regime). A discrete-time kMC model—with constant per-trial probability, open boundaries, and 4-neighbor coalescence—reproduces the measured time series of domain count and areal coverage using a single onset shift per condition. Quantitatively, the kMC-extracted onset times \(t_{0}\) (28.5~s and 38~s for Sets~1 and~2, respectively) match the experimentally observed no-growth thresholds (\(\sim\)30~s and \(\sim\)35~s), reinforcing that plasma dose sets a practical activation threshold for through-hole formation.  Raman mapping further confirms that graphene remains interfacial beneath GaN after high-temperature growth. 

Practically, these results establish a scalable, sub-lithographic route to tune GaN nucleation via plasma-engineered graphene masks and provide a quantitative process window for ELOG/THE with 2D masks. While direct threading-dislocation (TD) quantification is outside the present scope, the demonstrated control of nucleation density suggests a pathway to tailor TD density in future work. The approach is relevant to GaN devices where low defect densities are critical (e.g., $\mu$-LED displays, vertical power devices, and HEMTs), and it invites follow-up studies coupling this mask engineering with direct TD metrology (XRD/TEM/CL) and wafer-scale integration.

\section{acknowledgements}
This work was supported by the National Research Foundation of Korea(NRF) grant funded by the Korea government (MSIT) (RS-2021-NR060087, RS-2023-00240724) and through Korea Basic Science Institute (National research Facilities and Equipment Center) grant (2021R1A6C101A437) funded by the Ministry of Education.


\begin{thebibliography}{19}%
\makeatletter
\providecommand \@ifxundefined [1]{%
 \@ifx{#1\undefined}
}%
\providecommand \@ifnum [1]{%
 \ifnum #1\expandafter \@firstoftwo
 \else \expandafter \@secondoftwo
 \fi
}%
\providecommand \@ifx [1]{%
 \ifx #1\expandafter \@firstoftwo
 \else \expandafter \@secondoftwo
 \fi
}%
\providecommand \natexlab [1]{#1}%
\providecommand \enquote  [1]{``#1''}%
\providecommand \bibnamefont  [1]{#1}%
\providecommand \bibfnamefont [1]{#1}%
\providecommand \citenamefont [1]{#1}%
\providecommand \href@noop [0]{\@secondoftwo}%
\providecommand \href [0]{\begingroup \@sanitize@url \@href}%
\providecommand \@href[1]{\@@startlink{#1}\@@href}%
\providecommand \@@href[1]{\endgroup#1\@@endlink}%
\providecommand \@sanitize@url [0]{\catcode `\\12\catcode `\$12\catcode
  `\&12\catcode `\#12\catcode `\^12\catcode `\_12\catcode `\%12\relax}%
\providecommand \@@startlink[1]{}%
\providecommand \@@endlink[0]{}%
\providecommand \url  [0]{\begingroup\@sanitize@url \@url }%
\providecommand \@url [1]{\endgroup\@href {#1}{\urlprefix }}%
\providecommand \urlprefix  [0]{URL }%
\providecommand \Eprint [0]{\href }%
\providecommand \doibase [0]{http://dx.doi.org/}%
\providecommand \selectlanguage [0]{\@gobble}%
\providecommand \bibinfo  [0]{\@secondoftwo}%
\providecommand \bibfield  [0]{\@secondoftwo}%
\providecommand \translation [1]{[#1]}%
\providecommand \BibitemOpen [0]{}%
\providecommand \bibitemStop [0]{}%
\providecommand \bibitemNoStop [0]{.\EOS\space}%
\providecommand \EOS [0]{\spacefactor3000\relax}%
\providecommand \BibitemShut  [1]{\csname bibitem#1\endcsname}%
\let\auto@bib@innerbib\@empty
\bibitem [{\citenamefont {Amano}\ \emph {et~al.}(1986)\citenamefont {Amano},
  \citenamefont {Sawaki}, \citenamefont {Akasaki},\ and\ \citenamefont
  {Toyoda}}]{Amano-APL-48-353}%
  \BibitemOpen
  \bibfield  {author} {\bibinfo {author} {\bibfnamefont {H.}~\bibnamefont
  {Amano}}, \bibinfo {author} {\bibfnamefont {N.}~\bibnamefont {Sawaki}},
  \bibinfo {author} {\bibfnamefont {I.}~\bibnamefont {Akasaki}}, \ and\
  \bibinfo {author} {\bibfnamefont {Y.}~\bibnamefont {Toyoda}},\ }\href@noop {}
  {\bibfield  {journal} {\bibinfo  {journal} {Appl. Phys. Lett.}\ }\textbf
  {\bibinfo {volume} {48}},\ \bibinfo {pages} {353} (\bibinfo {year}
  {1986})}\BibitemShut {NoStop}%
\bibitem [{\citenamefont {Mathis}\ \emph {et~al.}(2001)\citenamefont {Mathis},
  \citenamefont {Romanov}, \citenamefont {Chen}, \citenamefont {Beltz},
  \citenamefont {Pompe},\ and\ \citenamefont {Speck}}]{Mathis-JCG-231-371}%
  \BibitemOpen
  \bibfield  {author} {\bibinfo {author} {\bibfnamefont {S.}~\bibnamefont
  {Mathis}}, \bibinfo {author} {\bibfnamefont {A.}~\bibnamefont {Romanov}},
  \bibinfo {author} {\bibfnamefont {L.}~\bibnamefont {Chen}}, \bibinfo {author}
  {\bibfnamefont {G.}~\bibnamefont {Beltz}}, \bibinfo {author} {\bibfnamefont
  {W.}~\bibnamefont {Pompe}}, \ and\ \bibinfo {author} {\bibfnamefont
  {J.}~\bibnamefont {Speck}},\ }\href@noop {} {\bibfield  {journal} {\bibinfo
  {journal} {J. Cryst. Growth}\ }\textbf {\bibinfo {volume} {231}},\ \bibinfo
  {pages} {371} (\bibinfo {year} {2001})}\BibitemShut {NoStop}%
\bibitem [{\citenamefont {Moram}\ \emph {et~al.}(2009)\citenamefont {Moram},
  \citenamefont {Ghedia}, \citenamefont {Rao}, \citenamefont {Barnard},
  \citenamefont {Zhang}, \citenamefont {Kappers},\ and\ \citenamefont
  {Humphreys}}]{Moram-JAP-106-073513}%
  \BibitemOpen
  \bibfield  {author} {\bibinfo {author} {\bibfnamefont {M.}~\bibnamefont
  {Moram}}, \bibinfo {author} {\bibfnamefont {C.}~\bibnamefont {Ghedia}},
  \bibinfo {author} {\bibfnamefont {D.}~\bibnamefont {Rao}}, \bibinfo {author}
  {\bibfnamefont {J.}~\bibnamefont {Barnard}}, \bibinfo {author} {\bibfnamefont
  {Y.}~\bibnamefont {Zhang}}, \bibinfo {author} {\bibfnamefont
  {M.}~\bibnamefont {Kappers}}, \ and\ \bibinfo {author} {\bibfnamefont
  {C.}~\bibnamefont {Humphreys}},\ }\href@noop {} {\bibfield  {journal}
  {\bibinfo  {journal} {J. Appl. Phys.}\ }\textbf {\bibinfo {volume} {106}},\
  \bibinfo {pages} {073513} (\bibinfo {year} {2009})}\BibitemShut {NoStop}%
\bibitem [{\citenamefont {Kaganer}\ \emph {et~al.}(2005)\citenamefont
  {Kaganer}, \citenamefont {Brandt}, \citenamefont {Trampert},\ and\
  \citenamefont {Ploog}}]{Kaganer-PRB-72-045423}%
  \BibitemOpen
  \bibfield  {author} {\bibinfo {author} {\bibfnamefont {V.~M.}\ \bibnamefont
  {Kaganer}}, \bibinfo {author} {\bibfnamefont {O.}~\bibnamefont {Brandt}},
  \bibinfo {author} {\bibfnamefont {A.}~\bibnamefont {Trampert}}, \ and\
  \bibinfo {author} {\bibfnamefont {K.}~\bibnamefont {Ploog}},\ }\href@noop {}
  {\bibfield  {journal} {\bibinfo  {journal} {Phys. Rev. B.}\ }\textbf
  {\bibinfo {volume} {72}},\ \bibinfo {pages} {045423} (\bibinfo {year}
  {2005})}\BibitemShut {NoStop}%
\bibitem [{\citenamefont {Kim}\ \emph {et~al.}(2020)\citenamefont {Kim},
  \citenamefont {Jang}, \citenamefont {Lee}, \citenamefont {Kim}, \citenamefont
  {Jang}, \citenamefont {Yoon},\ and\ \citenamefont {Kim}}]{Kim-CGD-20-6198}%
  \BibitemOpen
  \bibfield  {author} {\bibinfo {author} {\bibfnamefont {D.}~\bibnamefont
  {Kim}}, \bibinfo {author} {\bibfnamefont {D.}~\bibnamefont {Jang}}, \bibinfo
  {author} {\bibfnamefont {H.}~\bibnamefont {Lee}}, \bibinfo {author}
  {\bibfnamefont {J.}~\bibnamefont {Kim}}, \bibinfo {author} {\bibfnamefont
  {Y.}~\bibnamefont {Jang}}, \bibinfo {author} {\bibfnamefont {S.}~\bibnamefont
  {Yoon}}, \ and\ \bibinfo {author} {\bibfnamefont {C.}~\bibnamefont {Kim}},\
  }\href@noop {} {\bibfield  {journal} {\bibinfo  {journal} {Crystal Growth \&
  Design}\ }\textbf {\bibinfo {volume} {20}},\ \bibinfo {pages} {6198}
  (\bibinfo {year} {2020})}\BibitemShut {NoStop}%
\bibitem [{\citenamefont {K{\"u}pers}\ \emph {et~al.}(2017)\citenamefont
  {K{\"u}pers}, \citenamefont {Tahraoui}, \citenamefont {Lewis}, \citenamefont
  {Rauwerdink}, \citenamefont {Matalla}, \citenamefont {Kr{\"u}ger},
  \citenamefont {Bastiman}, \citenamefont {Riechert},\ and\ \citenamefont
  {Geelhaar}}]{Kupers-SST-32-115003}%
  \BibitemOpen
  \bibfield  {author} {\bibinfo {author} {\bibfnamefont {H.}~\bibnamefont
  {K{\"u}pers}}, \bibinfo {author} {\bibfnamefont {A.}~\bibnamefont
  {Tahraoui}}, \bibinfo {author} {\bibfnamefont {R.~B.}\ \bibnamefont {Lewis}},
  \bibinfo {author} {\bibfnamefont {S.}~\bibnamefont {Rauwerdink}}, \bibinfo
  {author} {\bibfnamefont {M.}~\bibnamefont {Matalla}}, \bibinfo {author}
  {\bibfnamefont {O.}~\bibnamefont {Kr{\"u}ger}}, \bibinfo {author}
  {\bibfnamefont {F.}~\bibnamefont {Bastiman}}, \bibinfo {author}
  {\bibfnamefont {H.}~\bibnamefont {Riechert}}, \ and\ \bibinfo {author}
  {\bibfnamefont {L.}~\bibnamefont {Geelhaar}},\ }\href@noop {} {\bibfield
  {journal} {\bibinfo  {journal} {Semiconductor Science and Technology}\
  }\textbf {\bibinfo {volume} {32}},\ \bibinfo {pages} {115003} (\bibinfo
  {year} {2017})}\BibitemShut {NoStop}%
\bibitem [{\citenamefont {Lin}\ \emph {et~al.}(2017)\citenamefont {Lin},
  \citenamefont {Ou}, \citenamefont {Aagesen}, \citenamefont {Jensen},
  \citenamefont {Herstr{\o}m},\ and\ \citenamefont {Ou}}]{Lin-MM-8-13}%
  \BibitemOpen
  \bibfield  {author} {\bibinfo {author} {\bibfnamefont {L.}~\bibnamefont
  {Lin}}, \bibinfo {author} {\bibfnamefont {Y.}~\bibnamefont {Ou}}, \bibinfo
  {author} {\bibfnamefont {M.}~\bibnamefont {Aagesen}}, \bibinfo {author}
  {\bibfnamefont {F.}~\bibnamefont {Jensen}}, \bibinfo {author} {\bibfnamefont
  {B.}~\bibnamefont {Herstr{\o}m}}, \ and\ \bibinfo {author} {\bibfnamefont
  {H.}~\bibnamefont {Ou}},\ }\href@noop {} {\bibfield  {journal} {\bibinfo
  {journal} {Micromachines}\ }\textbf {\bibinfo {volume} {8}},\ \bibinfo
  {pages} {13} (\bibinfo {year} {2017})}\BibitemShut {NoStop}%
\bibitem [{\citenamefont {Lee}\ \emph {et~al.}(2022)\citenamefont {Lee},
  \citenamefont {Kim}, \citenamefont {Jang}, \citenamefont {Jang},
  \citenamefont {Park},\ and\ \citenamefont {Kim}}]{Lee-CGD-22-6995}%
  \BibitemOpen
  \bibfield  {author} {\bibinfo {author} {\bibfnamefont {H.}~\bibnamefont
  {Lee}}, \bibinfo {author} {\bibfnamefont {M.}~\bibnamefont {Kim}}, \bibinfo
  {author} {\bibfnamefont {D.}~\bibnamefont {Jang}}, \bibinfo {author}
  {\bibfnamefont {S.}~\bibnamefont {Jang}}, \bibinfo {author} {\bibfnamefont
  {W.~I.}\ \bibnamefont {Park}}, \ and\ \bibinfo {author} {\bibfnamefont
  {C.}~\bibnamefont {Kim}},\ }\href@noop {} {\bibfield  {journal} {\bibinfo
  {journal} {Crystal Growth \& Design}\ }\textbf {\bibinfo {volume} {22}},\
  \bibinfo {pages} {6995} (\bibinfo {year} {2022})}\BibitemShut {NoStop}%
\bibitem [{\citenamefont {Jang}\ \emph {et~al.}(2023)\citenamefont {Jang},
  \citenamefont {Ahn}, \citenamefont {Lee}, \citenamefont {Lee}, \citenamefont
  {Lee}, \citenamefont {Kim}, \citenamefont {Kim}, \citenamefont {Park},
  \citenamefont {Kwon}, \citenamefont {Choi} \emph
  {et~al.}}]{Jang-AMI-10-2201406}%
  \BibitemOpen
  \bibfield  {author} {\bibinfo {author} {\bibfnamefont {D.}~\bibnamefont
  {Jang}}, \bibinfo {author} {\bibfnamefont {C.}~\bibnamefont {Ahn}}, \bibinfo
  {author} {\bibfnamefont {Y.}~\bibnamefont {Lee}}, \bibinfo {author}
  {\bibfnamefont {S.}~\bibnamefont {Lee}}, \bibinfo {author} {\bibfnamefont
  {H.}~\bibnamefont {Lee}}, \bibinfo {author} {\bibfnamefont {D.}~\bibnamefont
  {Kim}}, \bibinfo {author} {\bibfnamefont {Y.}~\bibnamefont {Kim}}, \bibinfo
  {author} {\bibfnamefont {J.-Y.}\ \bibnamefont {Park}}, \bibinfo {author}
  {\bibfnamefont {Y.-K.}\ \bibnamefont {Kwon}}, \bibinfo {author}
  {\bibfnamefont {J.}~\bibnamefont {Choi}},  \emph {et~al.},\ }\href@noop {}
  {\bibfield  {journal} {\bibinfo  {journal} {Advanced Materials Interfaces}\
  }\textbf {\bibinfo {volume} {10}},\ \bibinfo {pages} {2201406} (\bibinfo
  {year} {2023})}\BibitemShut {NoStop}%
\bibitem [{\citenamefont {Lee}\ \emph {et~al.}(2024)\citenamefont {Lee},
  \citenamefont {Lee}, \citenamefont {Choi}, \citenamefont {Kim},\ and\
  \citenamefont {Kwon}}]{Lee-AEM-26-2301654}%
  \BibitemOpen
  \bibfield  {author} {\bibinfo {author} {\bibfnamefont {Y.}~\bibnamefont
  {Lee}}, \bibinfo {author} {\bibfnamefont {S.}~\bibnamefont {Lee}}, \bibinfo
  {author} {\bibfnamefont {J.}~\bibnamefont {Choi}}, \bibinfo {author}
  {\bibfnamefont {C.}~\bibnamefont {Kim}}, \ and\ \bibinfo {author}
  {\bibfnamefont {Y.-K.}\ \bibnamefont {Kwon}},\ }\href@noop {} {\bibfield
  {journal} {\bibinfo  {journal} {Advanced Engineering Materials}\ }\textbf
  {\bibinfo {volume} {26}},\ \bibinfo {pages} {2301654} (\bibinfo {year}
  {2024})}\BibitemShut {NoStop}%
\bibitem [{\citenamefont {Beak}\ \emph {et~al.}(2025)\citenamefont {Beak},
  \citenamefont {Dong}, \citenamefont {Choi}, \citenamefont {Yang},
  \citenamefont {Lim},\ and\ \citenamefont {Kim}}]{Beak-arXiv}%
  \BibitemOpen
  \bibfield  {author} {\bibinfo {author} {\bibfnamefont {G.}~\bibnamefont
  {Beak}}, \bibinfo {author} {\bibfnamefont {C.}~\bibnamefont {Dong}}, \bibinfo
  {author} {\bibfnamefont {M.}~\bibnamefont {Choi}}, \bibinfo {author}
  {\bibfnamefont {J.}~\bibnamefont {Yang}}, \bibinfo {author} {\bibfnamefont
  {J.}~\bibnamefont {Lim}}, \ and\ \bibinfo {author} {\bibfnamefont
  {C.}~\bibnamefont {Kim}},\ }\href@noop {} {\bibfield  {journal} {\bibinfo
  {journal} {arXiv preprint arXiv:2505.04454}\ } (\bibinfo {year}
  {2025})}\BibitemShut {NoStop}%
\bibitem [{\citenamefont {Ha}\ \emph {et~al.}(2025)\citenamefont {Ha},
  \citenamefont {Choi}, \citenamefont {Yang},\ and\ \citenamefont
  {Kim}}]{Ha-arXiv}%
  \BibitemOpen
  \bibfield  {author} {\bibinfo {author} {\bibfnamefont {J.}~\bibnamefont
  {Ha}}, \bibinfo {author} {\bibfnamefont {M.}~\bibnamefont {Choi}}, \bibinfo
  {author} {\bibfnamefont {J.}~\bibnamefont {Yang}}, \ and\ \bibinfo {author}
  {\bibfnamefont {C.}~\bibnamefont {Kim}},\ }\href@noop {} {\bibfield
  {journal} {\bibinfo  {journal} {arXiv preprint arXiv:2505.11045}\ } (\bibinfo
  {year} {2025})}\BibitemShut {NoStop}%
\bibitem [{\citenamefont {Chen}\ \emph {et~al.}(2023)\citenamefont {Chen},
  \citenamefont {Yang}, \citenamefont {Shi}, \citenamefont {Yi}, \citenamefont
  {Wang}, \citenamefont {Li},\ and\ \citenamefont {Liu}}]{Chen-AM-35-2211075}%
  \BibitemOpen
  \bibfield  {author} {\bibinfo {author} {\bibfnamefont {Q.}~\bibnamefont
  {Chen}}, \bibinfo {author} {\bibfnamefont {K.}~\bibnamefont {Yang}}, \bibinfo
  {author} {\bibfnamefont {B.}~\bibnamefont {Shi}}, \bibinfo {author}
  {\bibfnamefont {X.}~\bibnamefont {Yi}}, \bibinfo {author} {\bibfnamefont
  {J.}~\bibnamefont {Wang}}, \bibinfo {author} {\bibfnamefont {J.}~\bibnamefont
  {Li}}, \ and\ \bibinfo {author} {\bibfnamefont {Z.}~\bibnamefont {Liu}},\
  }\href@noop {} {\bibfield  {journal} {\bibinfo  {journal} {Advanced
  Materials}\ }\textbf {\bibinfo {volume} {35}},\ \bibinfo {pages} {2211075}
  (\bibinfo {year} {2023})}\BibitemShut {NoStop}%
\bibitem [{\citenamefont {Lee}\ \emph {et~al.}(2023)\citenamefont {Lee},
  \citenamefont {Abbas}, \citenamefont {Yoo}, \citenamefont {Lee},
  \citenamefont {Fabunmi}, \citenamefont {Lee}, \citenamefont {Kim},
  \citenamefont {Kim}, \citenamefont {Jang}, \citenamefont {Lee} \emph
  {et~al.}}]{Lee-NL-23-11578}%
  \BibitemOpen
  \bibfield  {author} {\bibinfo {author} {\bibfnamefont {S.}~\bibnamefont
  {Lee}}, \bibinfo {author} {\bibfnamefont {M.~S.}\ \bibnamefont {Abbas}},
  \bibinfo {author} {\bibfnamefont {D.}~\bibnamefont {Yoo}}, \bibinfo {author}
  {\bibfnamefont {K.}~\bibnamefont {Lee}}, \bibinfo {author} {\bibfnamefont
  {T.~G.}\ \bibnamefont {Fabunmi}}, \bibinfo {author} {\bibfnamefont
  {E.}~\bibnamefont {Lee}}, \bibinfo {author} {\bibfnamefont {H.~I.}\
  \bibnamefont {Kim}}, \bibinfo {author} {\bibfnamefont {I.}~\bibnamefont
  {Kim}}, \bibinfo {author} {\bibfnamefont {D.}~\bibnamefont {Jang}}, \bibinfo
  {author} {\bibfnamefont {S.}~\bibnamefont {Lee}},  \emph {et~al.},\
  }\href@noop {} {\bibfield  {journal} {\bibinfo  {journal} {Nano Letters}\
  }\textbf {\bibinfo {volume} {23}},\ \bibinfo {pages} {11578} (\bibinfo {year}
  {2023})}\BibitemShut {NoStop}%
\bibitem [{\citenamefont {Feng}\ \emph {et~al.}(2019)\citenamefont {Feng},
  \citenamefont {Yang}, \citenamefont {Zhang}, \citenamefont {Kang},
  \citenamefont {Zhang}, \citenamefont {Liu}, \citenamefont {Li}, \citenamefont
  {Shen}, \citenamefont {Liu}, \citenamefont {Wang} \emph
  {et~al.}}]{Feng-AFM-29-1905056}%
  \BibitemOpen
  \bibfield  {author} {\bibinfo {author} {\bibfnamefont {Y.}~\bibnamefont
  {Feng}}, \bibinfo {author} {\bibfnamefont {X.}~\bibnamefont {Yang}}, \bibinfo
  {author} {\bibfnamefont {Z.}~\bibnamefont {Zhang}}, \bibinfo {author}
  {\bibfnamefont {D.}~\bibnamefont {Kang}}, \bibinfo {author} {\bibfnamefont
  {J.}~\bibnamefont {Zhang}}, \bibinfo {author} {\bibfnamefont
  {K.}~\bibnamefont {Liu}}, \bibinfo {author} {\bibfnamefont {X.}~\bibnamefont
  {Li}}, \bibinfo {author} {\bibfnamefont {J.}~\bibnamefont {Shen}}, \bibinfo
  {author} {\bibfnamefont {F.}~\bibnamefont {Liu}}, \bibinfo {author}
  {\bibfnamefont {T.}~\bibnamefont {Wang}},  \emph {et~al.},\ }\href@noop {}
  {\bibfield  {journal} {\bibinfo  {journal} {Adv. Funct. Mater.}\ }\textbf
  {\bibinfo {volume} {29}},\ \bibinfo {pages} {1905056} (\bibinfo {year}
  {2019})}\BibitemShut {NoStop}%
\bibitem [{\citenamefont {Tao}\ \emph {et~al.}(2024)\citenamefont {Tao},
  \citenamefont {Xu}, \citenamefont {Li}, \citenamefont {Cai}, \citenamefont
  {Wang}, \citenamefont {Wang}, \citenamefont {Cao},\ and\ \citenamefont
  {Xu}}]{Tao-JJAP-63-025503}%
  \BibitemOpen
  \bibfield  {author} {\bibinfo {author} {\bibfnamefont {J.}~\bibnamefont
  {Tao}}, \bibinfo {author} {\bibfnamefont {Y.}~\bibnamefont {Xu}}, \bibinfo
  {author} {\bibfnamefont {J.}~\bibnamefont {Li}}, \bibinfo {author}
  {\bibfnamefont {X.}~\bibnamefont {Cai}}, \bibinfo {author} {\bibfnamefont
  {Y.}~\bibnamefont {Wang}}, \bibinfo {author} {\bibfnamefont {G.}~\bibnamefont
  {Wang}}, \bibinfo {author} {\bibfnamefont {B.}~\bibnamefont {Cao}}, \ and\
  \bibinfo {author} {\bibfnamefont {K.}~\bibnamefont {Xu}},\ }\href@noop {}
  {\bibfield  {journal} {\bibinfo  {journal} {Jpn. J. Appl. Phys.}\ }\textbf
  {\bibinfo {volume} {63}},\ \bibinfo {pages} {025503} (\bibinfo {year}
  {2024})}\BibitemShut {NoStop}%
\bibitem [{\citenamefont {Liu}\ \emph {et~al.}(2008)\citenamefont {Liu},
  \citenamefont {Ryu}, \citenamefont {Tomasik}, \citenamefont {Stolyarova},
  \citenamefont {Jung}, \citenamefont {Hybertsen}, \citenamefont {Steigerwald},
  \citenamefont {Brus},\ and\ \citenamefont {Flynn}}]{Liu-NL-8-1965}%
  \BibitemOpen
  \bibfield  {author} {\bibinfo {author} {\bibfnamefont {L.}~\bibnamefont
  {Liu}}, \bibinfo {author} {\bibfnamefont {S.}~\bibnamefont {Ryu}}, \bibinfo
  {author} {\bibfnamefont {M.~R.}\ \bibnamefont {Tomasik}}, \bibinfo {author}
  {\bibfnamefont {E.}~\bibnamefont {Stolyarova}}, \bibinfo {author}
  {\bibfnamefont {N.}~\bibnamefont {Jung}}, \bibinfo {author} {\bibfnamefont
  {M.~S.}\ \bibnamefont {Hybertsen}}, \bibinfo {author} {\bibfnamefont {M.~L.}\
  \bibnamefont {Steigerwald}}, \bibinfo {author} {\bibfnamefont {L.~E.}\
  \bibnamefont {Brus}}, \ and\ \bibinfo {author} {\bibfnamefont {G.~W.}\
  \bibnamefont {Flynn}},\ }\href@noop {} {\bibfield  {journal} {\bibinfo
  {journal} {Nano Lett.}\ }\textbf {\bibinfo {volume} {8}},\ \bibinfo {pages}
  {1965} (\bibinfo {year} {2008})}\BibitemShut {NoStop}%
\bibitem [{\citenamefont {Childres}\ \emph {et~al.}(2011)\citenamefont
  {Childres}, \citenamefont {Jauregui}, \citenamefont {Tian},\ and\
  \citenamefont {Chen}}]{Childres-NJP-13-025008}%
  \BibitemOpen
  \bibfield  {author} {\bibinfo {author} {\bibfnamefont {I.}~\bibnamefont
  {Childres}}, \bibinfo {author} {\bibfnamefont {L.~A.}\ \bibnamefont
  {Jauregui}}, \bibinfo {author} {\bibfnamefont {J.}~\bibnamefont {Tian}}, \
  and\ \bibinfo {author} {\bibfnamefont {Y.~P.}\ \bibnamefont {Chen}},\
  }\href@noop {} {\bibfield  {journal} {\bibinfo  {journal} {New J. Phys.}\
  }\textbf {\bibinfo {volume} {13}},\ \bibinfo {pages} {025008} (\bibinfo
  {year} {2011})}\BibitemShut {NoStop}%
\bibitem [{\citenamefont {Zhang}\ and\ \citenamefont
  {Gan}(2013)}]{Zhang-ASS-285-211}%
  \BibitemOpen
  \bibfield  {author} {\bibinfo {author} {\bibfnamefont {D.}~\bibnamefont
  {Zhang}}\ and\ \bibinfo {author} {\bibfnamefont {Y.}~\bibnamefont {Gan}},\
  }\href@noop {} {\bibfield  {journal} {\bibinfo  {journal} {Appl. Surf. Sci.}\
  }\textbf {\bibinfo {volume} {285}},\ \bibinfo {pages} {211} (\bibinfo {year}
  {2013})}\BibitemShut {NoStop}%
\end{thebibliography}

%


\end{document}


\preprint{}

\preprint{}

\title{{\normalsize{Supplementary Information for}} \\
Controlling GaN nucleation via O\textsubscript{2}-plasma-perforated graphene masks on c-plane sapphire}

\author{Su Young An}
\affiliation{Department of Physics, Kyung Hee University, Seoul 02447, Republic of Korea}

\author{Chinkyo Kim$^{*,}$}
\thanks{email: ckim@khu.ac.kr}
\affiliation{Department of Physics, Kyung Hee University, Seoul 02447, Republic of Korea}
\affiliation{Department of Information Display, Kyung Hee University, Seoul 02447, Republic of Korea}

\maketitle
\thispagestyle{empty}
\newpage

\begin{figure}
\includegraphics[width=1.0\columnwidth]{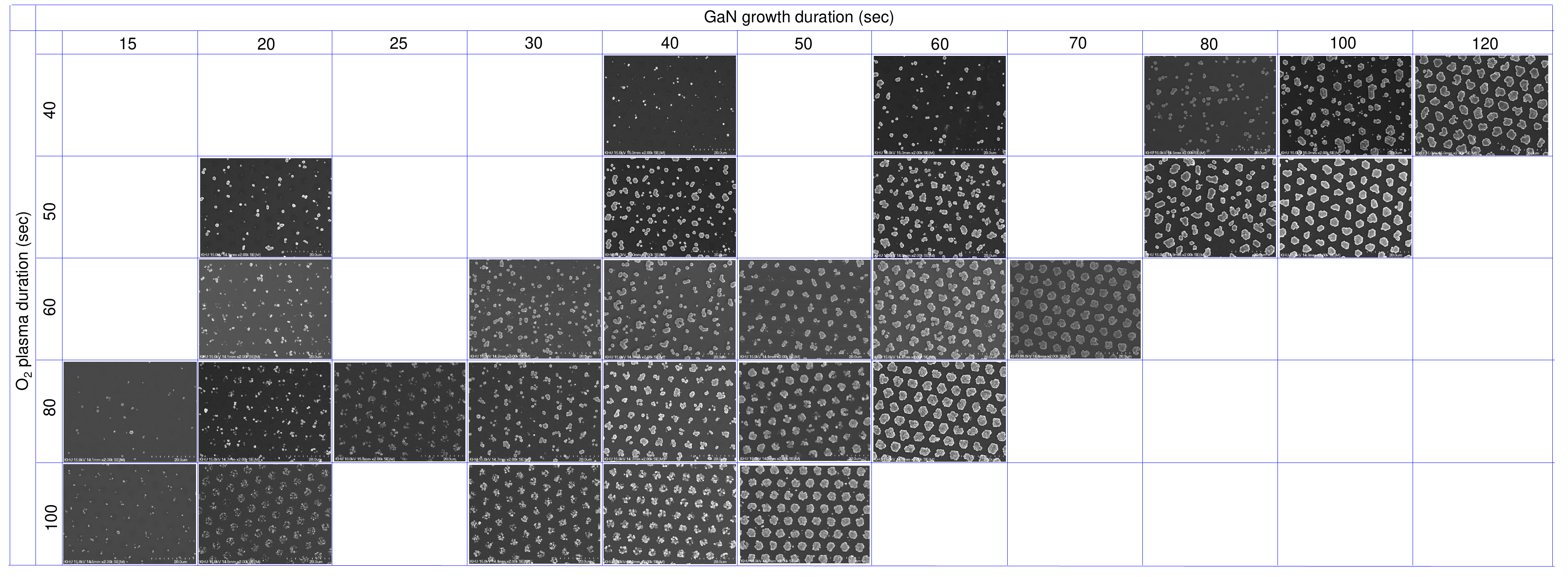}
\caption{SEM images of GaN domains grown on graphene/$c$-plane sapphire, with O$_2$-plasma exposure times and growth durations varied as indicated.}
\label{SEM_GaN_on_O2-etched_graphene_set2}
\end{figure}

\begin{figure}
\includegraphics[width=0.7\columnwidth]{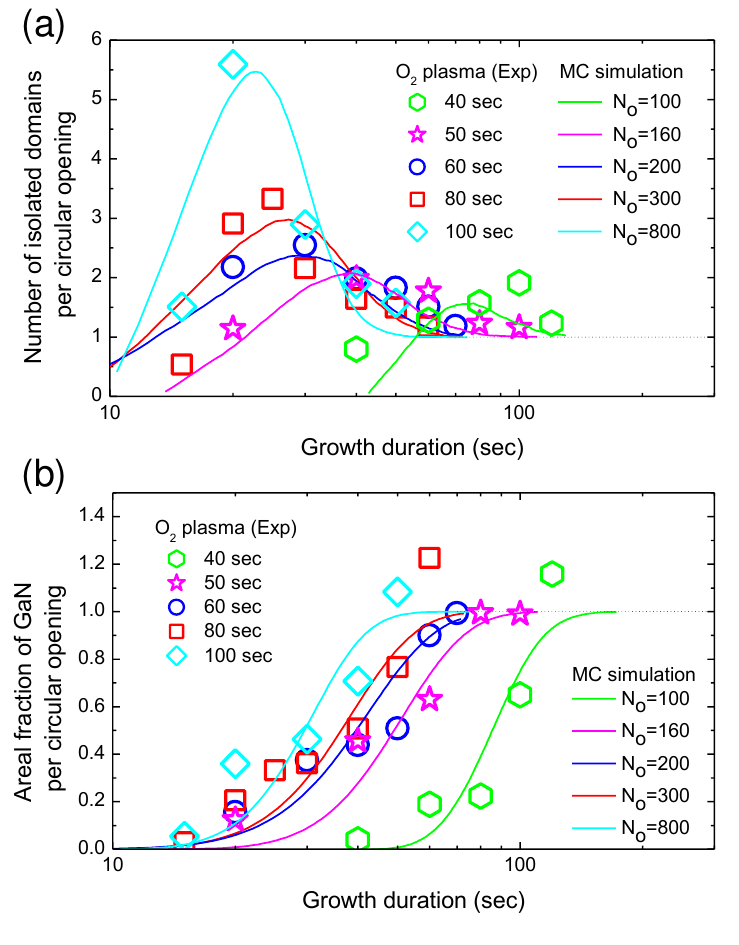}
\caption{Time evolution of the number density and areal fraction of GaN domains for Set~2. 
$N_{\mathrm t}=1\times10^{4}$, $P_{0}=5\times10^{-4}$, $N_{\mathrm r}=100$–$300$, $N_{\mathrm a}=2{,}000$.}
\label{Time-evolution-of-GaN-domains-set2}
\end{figure}


\newpage

\makeatletter
\renewcommand{\bibsection}{}
\makeatother


%